\documentclass[aps,prb,twocolumn,showpacs,floatfix,longbibliography,superscriptaddress,10pt]{revtex4-1}
\usepackage{graphicx,amsfonts,amssymb,amsmath}
\usepackage{textcomp}
\usepackage{esint}
\usepackage[colorlinks,linktocpage,bookmarks=false,citecolor=blue,linkcolor=red,urlcolor=blue]{hyperref}
\usepackage[T1]{fontenc}
\usepackage[dvipsnames]{xcolor}
\usepackage[english]{babel}
\usepackage{afterpage}

\allowdisplaybreaks

\newcommand{\red}[1]{\textcolor{red}{#1}}

\def\Jperp{J_\perp} 
\def\Tkt{T_{\rm KT}}
\def\qperp{q_\perp}
\def\pperp{p_\perp} 
 
\begin{document}

\graphicspath{{./figures_submit/}}
\title{Kosterlitz-Thouless signatures in the low-temperature phase of layered three-dimensional systems}

\author{Adam Ran\c{c}on}
\affiliation{Laboratoire de Physique des Lasers, Atomes et Mol\'ecules, Universit\'e Lille 1, 59655 Villeneuve d'Ascq cedex, France} 
\author{Nicolas Dupuis}
\affiliation{Laboratoire de Physique Th\'eorique de la Mati\`ere Condens\'ee,
CNRS UMR 7600, UPMC-Sorbonne Universit\'es, 4 Place Jussieu, 
75252 Paris Cedex 05, France}

\date{December 12, 2017} 

\begin{abstract}  
We study the  quasi-two-dimensional quantum O(2) model, a quantum generalization of the Lawrence-Doniach model, within the nonperturbative renormalization-group approach and propose a generic phase diagram for layered three-dimensional systems with an O(2)-symmetric order parameter. Below the transition temperature we identify a wide region of the phase diagram where the renormalization-group flow is quasi-two-dimensional for length scales smaller than a Josephson length $l_J$, leading to signatures of Kosterlitz-Thouless physics in the temperature dependence of physical observables. In particular the order parameter varies as a power law of the interplane coupling with an exponent which depends on the anomalous dimension (itself related to the stiffness) of the strictly two-dimensional low-temperature Kosterlitz-Thouless phase.
\end{abstract}

\maketitle


\def\rhoeq{\hat\rho_{\rm eq}}

\newcommand{\marge}[1]{\marginpar{\scriptsize #1}}
\newcommand{\remarque}[1]{\marginpar{\scriptsize Remarque}{\it [#1]}}
\newcommand{\new}[1]{{\bf #1}}
\newlength{\textlarg}
\newcommand{\barre}[1]{%
   \settowidth{\textlarg}{#1}
   #1\hspace{-\textlarg}\rule[0.5ex]{\textlarg}{0.5pt}}
\newcommand{\barred}[1]{%
   \settowidth{\textlarg}{#1}
   \red{#1}\hspace{-\textlarg}\rule[0.5ex]{\textlarg}{0.5pt}}
\newcommand{\barblue}[1]{%
   \settowidth{\textlarg}{#1}
   \blue{#1}\hspace{-\textlarg}\rule[0.5ex]{\textlarg}{0.5pt}}

\newcommand{\beq}{\begin{equation}}
\newcommand{\eeq}{\end{equation}}
\newcommand{\bleq}{\begin{eqnarray}}
\newcommand{\eleq}{\end{eqnarray}} 
\newcommand{\bfig}{\begin{figure}}
\newcommand{\efig}{\end{figure}}
\newcommand{\bline}{\begin{multline}}
\newcommand{\eline}{\end{multline}}
\newcommand{\bremark}{\begin{quotation} \noindent \small }
\newcommand{\eremark}{\end{quotation}}
\newcommand{\llbrace}{\left\lbrace}
\newcommand{\rrbrace}{\right\rbrace}
\newcommand{\lbraket}{\left[}
\newcommand{\rbraket}{\right]}
\newcommand{\llangle}{\left\langle}
\newcommand{\rrangle}{\right\rangle} 

\newcommand{\Tr}{{\rm Tr}} 
\newcommand{\tr}{{\rm tr}} 
\newcommand{\sgn}{\,{\rm sgn}} 
\newcommand{\mean}[1]{\langle #1 \rangle}
\newcommand{\commu}[2]{[#1,#2]} 
\newcommand{\bra}[1]{\langle#1|}
\newcommand{\ket}[1]{|#1\rangle}
\newcommand{\braket}[2]{\langle #1|#2\rangle}
\newcommand{\dbraket}[3]{\langle #1|#2|#3\rangle}
\newcommand{\tens}[1]{\overleftrightarrow{#1}}  
\newcommand{\vac}{|{\rm vac}\rangle} 
\newcommand{\bravac}{\langle{\rm vac}|}
\newcommand{\const}{{\rm const}} 
\newcommand{\atanh}{\,{\rm atanh}}
\newcommand{\cotanh}{\,{\rm cotanh}}

\newcommand{\ie}{i.e.\xspace}
\newcommand{\iet}{i.e.}
\newcommand{\eg}{e.g.\xspace}
\newcommand{\cc}{{\rm c.c.}} 
\newcommand{\hc}{{\rm h.c.}} 
\newcommand{\etal}{{\it et al. }}

\newcommand{\jhatbf}{\hat {\textbf \jold}} 
\newcommand{\Jhatbf}{\hat {\textbf \J}} 
\newcommand{\jhat}{\hat {\jmath}} 
\newcommand{\Jhat}{\hat {J}} 
\newcommand{\jbf}{\textbf j}
\newcommand{\Jbf}{\textbf J}

\def\chibf{\boldsymbol{\chi}}
\def\down{\downarrow}
\def\eps{\epsilon}
\def\gam{\gamma} 
\def\phibf{\boldsymbol{\phi}}
\def\varphibf{\boldsymbol{\varphi}}
\def\varphibfs{\boldsymbol{\varphi}_<}
\def\varphibfl{\boldsymbol{\varphi}_>}
\def\varphis{\varphi_{<}}
\def\varphil{\varphi_{>}}
\def\psibf{\boldsymbol{\psi}}
\def\thetabf{\boldsymbol{\theta}}
\def\Ome{\Omega}
\def\omeD{{\omega_D}} 
\def\bfOme{\boldsymbol{\Omega}} 
\def\Omebf{\boldsymbol{\Omega}} 
\def\lamb{\lambda}
\def\Lamb{\Lambda}
\def\sig{\sigma}
\def\Sig{\Sigma}
\def\sigp{{\sigma'}} 
\def\bfsig{\boldsymbol{\sigma}} 
\def\sigbf{\boldsymbol{\sigma}} 
\def\bfSig{\boldsymbol{\Sigma}} 
\def\The{\Theta} 
\def\up{\uparrow}

\def\epsk{\epsilon_{\bf k}} 
\def\xik{\xi_{\bf k}} 
\def\txik{\tilde\xi_{\bf k}} 
\def\xip{\xi_{\bf p}} 
\def\xiq{\xi_{\bf q}} 
\def\xikq{\xi_{{\bf k}+{\bf q}}} 
\def\Ek{E_{\bf k}} 
\def\Ep{E_{\bf p}}
\def\Eq{E_{\bf q}}
\def\Heff{\hat H_{\rm eff}}
\def\Hem{\hat H_{\rm em}}
\def\Hint{\hat H_{\rm int}}
\def\Hloc{\hat H_{\rm loc}}
\def\HMF{\hat H_{\rm MF}}
\def\Sem{S_{\rm em}}
\def\SMF{S_{\rm MF}} 
\def\SHF{S_{\rm HF}} 
\def\SRPA{S_{\rm RPA}} 
\def\Sint{S_{\rm int}} 
\def\Sloc{S_{\rm loc}}
\def\TN{T_{\rm N}} 
\def\TNHF{T^{\rm HF}_{\rm N}} 
\def\Zloc{Z_{\rm loc}} 
\def\ZMF{Z_{\rm MF}} 
\def\ZHF{Z_{\rm HF}} 
\def\ZRPA{Z_{\rm RPA}} 
\def\RPA{{\rm RPA}}
\def\loc{{\rm loc}} 
\def\pp{{\rm pp}}
\def\ph{{\rm ph}} 
\def\ch{{\rm ch}}
\def\sp{{\rm sp}} 
\def\qtf{q_{\rm TF}}
\def\epstf{\eps^{}_{\rm TF}} 
\def\epsrpa{\eps^{}_{\rm RPA}} 
\def\chinnzpp{\chi_{nn}^{0}{}\!\!\!''}

\def\half{\frac{1}{2}}
\def\dhalf{\dfrac{1}{2}}
\def\third{\frac{1}{3}} 
\def\quarter{\frac{1}{4}}

\def\qr{{\bf q}\cdot{\bf r}}
\def\wt{\omega t} 

\def\a{{\bf a}}
\def\b{{\bf b}}
\newcommand{\cv}{{\bf c}} 
\def\e{{\bf e}}
\def\f{{\bf f}}
\def\g{{\bf g}}
\def\h{{\bf h}}
\def\jold{\char"11}
\def\j{{\bf j}}
\def\k{{\bf k}}
\def\l{{\bf l}}
\def\m{{\bf m}}
\def\n{{\bf n}} 
\def\p{{\bf p}} 
\def\q{{\bf q}}
\def\r{{\bf r}}
\def\t{{\bf t}}
\def\u{{\bf u}}
\newcommand{\vv}{{\bf v}}
\def\x{{\bf x}}
\def\y{{\bf y}} 
\def\z{{\bf z}} 
\def\A{{\bf A}}
\def\B{{\bf B}}
\def\D{{\bf D}} 
\def\E{{\bf E}} 
\def\F{{\bf F}} 
\def\H{{\bf H}}  
\def\J{{\bf J}}
\def\K{{\bf K}} 

\def\G{{\bf G}}
\def\L{{\bf L}}
\def\M{{\bf M}}  
\def\O{{\bf O}} 
\def\P{{\bf P}} 
\def\Q{{\bf Q}} 
\def\R{{\bf R}}
\def\S{{\bf S}}
\def\U{{\bf U}} 
\def\X{{\bf X}} 
\def\Y{{\bf Y}} 
\def\epsbf{\boldsymbol{\epsilon}}
\def\betabf{\boldsymbol{\beta}}
\def\mubf{\boldsymbol{\mu}}
\def\nablabf{\boldsymbol{\nabla}}
\def\rhobf{\boldsymbol{\rho}}
\def\sigmabf{\boldsymbol{\sigma}} 
\def\Pibf{\boldsymbol{\Pi}}
\def\pibf{\boldsymbol{\pi}}

\def\para{\parallel}
\def\kpara{{k_\parallel}}
\def\kperp{{k_\perp}} 
\def\kperpp{{k_\perp'}} 
\def\qperp{{q_\perp}} 
\def\tperp{{t_\perp}} 

\def\w{\omega}
\def\wn{\omega_n}
\def\wm{\omega_m}
\def\wnu{\omega_\nu}
\def\wp{\omega_p} 
\def\dmu{{\partial_\mu}}
\def\dnu{{\partial_\nu}}
\def\dl{{\partial_l}}  
\def\dt{\partial_t} 
\def\tdt{\tilde\partial_t}
\def\dk{\partial_k}
\def\tdk{\tilde\partial_k}
\def\dx{\partial_x}
\def\dy{\partial_y} 
\def\dtau{{\partial_\tau}}  
\def\det{{\rm det}} 
\def\Pf{{\rm Pf}}
\def\diag{{\rm diag}}

\def\dsum{\displaystyle \sum}
\def\dint{\displaystyle \int} 
\def\intt{\int_{-\infty}^\infty dt} 
\def\inttp{\int_{-\infty}^\infty dt'} 
\def\intk{\int_{\bf k}} 
\def\intkd{\int \frac{d^dk}{(2\pi)^d}}
\def\intq{\int_{\bf q}} 
\def\intr{\int d^dr}  
\def\dintr{\displaystyle \int d^dr} 
\def\intrp{\int d^dr'}
\def\dinttau{\displaystyle \int_0^\beta d\tau}
\def\dinttaup{\displaystyle \int_0^\beta d\tau'}
\def\inttau{\int_0^\beta d\tau}
\def\inttaup{\int_0^\beta d\tau'}
\def\intx{\int d^{d+1}x} 
\def\inttaur{\int_0^\beta d\tau \int d^dr}
\def\intinf{\int_{-\infty}^\infty}
\def\dinttaur{\displaystyle \int_0^\beta d\tau \int d^dr}
\def\dintinf{\displaystyle \int_{-\infty}^\infty}
\def\intw{\int_{-\infty}^\infty \frac{d\w}{2\pi}}
\def\sumr{\sum_{\bf r}} 

\def\calA{{\cal A}}
\def\calAbf{\bm{{\cal A}}}
\def\calB{{\cal B}} 
\def\calC{{\cal C}} 
\def\dt{\partial_t}
\def\calD{{\cal D}}
\def\calE{{\cal E}}
\def\calF{{\cal F}} 
\def\calFbf{\bm{{\cal F}}}
\def\calG{{\cal G}}
\def\calH{{\cal H}}
\def\calI{{\cal I}}
\def\calJ{{\cal J}}
\def\calK{{\cal K}}
\def\calL{{\cal L}} 
\def\calM{{\cal M}} 
\def\calN{{\cal N}}
\def\calO{{\cal O}}
\def\calP{{\cal P}}  
\def\calR{{\cal R}} 
\def\calS{{\cal S}}
\def\calT{{\cal T}}
\def\calU{{\cal U}}
\def\calX{{\cal X}} 
\def\calY{{\cal Y}} 
\def\calZ{{\cal Z}} 

\def\calbfB{{\bf \cal B}}
\def\calbfF{{\bf \cal F}}

\def\tT{{\tilde T}}
\def\talpha{{\tilde\alpha}}
\def\tdelta{{\tilde\delta}}
\def\teta{{\tilde\eta}} 
\def\tlamb{{\tilde\lambda}}
\def\tmu{{\tilde\mu}}
\def\tphibf{{\tilde\phibf}}
\def\trho{{\tilde\rho}}
\def\tvarphibf{{\tilde\varphibf}} 
\def\tw{{\tilde\omega}}
\def\twn{{\tilde\omega_n}}
\def\twnu{{\tilde\omega_\nu}}

\def\asinh{{\rm asinh}} 

\section{Introduction}

The nature of second-order thermal phase transitions strongly depends on space dimension and number of components $N$ of the order parameter. In two dimensions the Mermin-Wagner theorem forbids any finite-temperature long-range order when $N\geq 2$. For $N=2$, there is nevertheless a transition at a finite temperature $\Tkt$ below which the order parameter correlation function decays algebraically.\cite{Berezinskii70,*Berezinskii71,Kosterlitz73,Kosterlitz74} The key role of topological defects in this transition was recognized by Kosterlitz and Thouless (KT) who formulated the transition as a vortex-antivortex-pair unbinding transition.\cite{Kosterlitz73,Kosterlitz74,Jose77,Ambegaokar80,Minnhagen87}
The KT transition differs from more conventional finite-temperature phase transitions in a number of aspects. It is not characterized by spontaneous symmetry breaking but the system exhibits a nonzero ``stiffness'' $\rho_s(T)$ for all temperatures $T < \Tkt$. Above the transition temperature $\Tkt$, one observes a standard disordered phase with exponentially decaying correlation functions. However, the correlation length $\xi$ does not diverge as a power law of $\tau=T-\Tkt$ but shows an essential singularity $\xi\sim\exp(\const/\sqrt{\tau})$. The transition is also characterized by a jump of the stiffness $\rho_s(T)$ which vanishes for $T>\Tkt$ and takes the universal value $2/\pi$ for $T\to\Tkt^-$.\cite{Nelson77a,Minnhagen81} 

Since many materials actually consist of weakly-coupled (rather than isolated) planes, one can wonder to what extent
finite-scale quasi-two-dimensional topological excitations may give observable signatures of KT physics in a layered three-dimensional (3D) system. This issue has raised a lot of interest in the 1990s in connection with high-temperature superconductivity. Recent advances in the physics of two-dimensional cold atomic gases\cite{Desbuquois12,Desbuquois14} may provide us with a fresh experimental insight into this problem. 

Most theoretical studies so far have relied on the layered 3D XY model,\cite{Berezinskii73,Pokrovskii74,Hikami80,Janke90,Minnhagen91,Schmidt92,Pierson92a,Pierson92b,Fischer93,Chattopadhyay94,Shenoy95,Pierson95,Friesen95} many properties of which being now well understood. The interplane coupling is a relevant (in the renormalization-group sense) perturbation so that the transition is 3D and belongs to the universality class of the 3D isotropic XY model (or 3D O(2) model).\cite{Pierson92a,Pierson92b,Fischer93} The transition temperature $T_c$ tends to $\Tkt$ as the interplane coupling vanishes\cite{Hikami80,Pierson92a,Pierson92b,Fischer93,Laflorencie12,Furuya16} with an upward shift $T_c-\Tkt$ which depends logarithmically on the anisotropy of the system.\cite{Hikami80,Pierson92a,Pierson92b} Even though the correlation length $\xi\sim (T-T_c)^{-\nu}$ in the parallel direction ultimately diverges for $T\to T_c$ as a power law with an exponent $\nu$ given by the critical exponent of the 3D XY universality class, there is a temperature range above $T_c$ where 2D KT scaling $\xi\sim \exp(\const/\sqrt{\tau})$, $\tau=T-\Tkt^{\rm eff}$, is observed with an effective KT temperature $\Tkt^{\rm eff}$ of the order of $T_c$.\cite{Shenoy95,Benfatto07} The jump in the stiffness, which characterizes the 2D KT transition, is replaced by a rapid (but continuous) suppression at a temperature of the order of $\Tkt$.\cite{Chattopadhyay94,Benfatto07} For $T<T_c$ and in the limit of small inter-layer coupling $\Jperp$, the order parameter depends on the interplane coupling $\Jperp$ as a power law with an exponent which depends on the anomalous dimension $\eta(T)=1/2\pi\rho_s(T)$ of the strictly 2D KT phase.\cite{Berezinskii73,Pokrovskii74,Hikami80}  

In this manuscript, we study layered systems with an O(2)-symmetric order parameter using a quasi-2D quantum O(2) model which can be seen as a quantum generalization of the Lawrence-Doniach model.\cite{Lawrence70,[{For a recent study of a related (although different) dimensional crossover in Bose systems confined within slab geometries, see }] Delfino17}
This model captures quantum fluctuations at low temperatures but describes otherwise qualitatively the same physics as the (classical) layered 3D XY model. Our approach is based on the nonperturbative renormalization-group approach (NPRG) and, contrary to most previous works, does not introduce vortices explicitly. While only a refined approximation scheme of the NPRG equations is able to capture all features of the KT transition and in particular the low-temperature line of fixed points,\cite{Jakubczyk14,Jakubczyk16,Jakubczyk17,[{See also }]Krieg17a,*Defenu17} we shall use a simple approximation\cite{Graeter95,Gersdorff01} which is easily numerically tractable even in the quasi-2D case. In this approximation, 
the KT transition is not captured {\it stricto sensu} since the correlation length is always finite. Nevertheless, below a ``transition'' temperature $\Tkt$ one finds a line of quasi-fixed points implying a very large correlation length. Thus, although our approach does not yield a low-temperature phase with an infinite correlation length, it nevertheless allows one to estimate the KT transition temperature and gives a satisfying description of algebraic (quasi-long-range) order in the low-temperature phase. (For various applications of this approach to the KT transition, see Refs.~\onlinecite{Krahl07,Machado10,Floerchinger09a,Rancon12b,Rancon13b,Rancon14a}.)

\begin{figure} 
\centerline{\includegraphics[width=8cm]{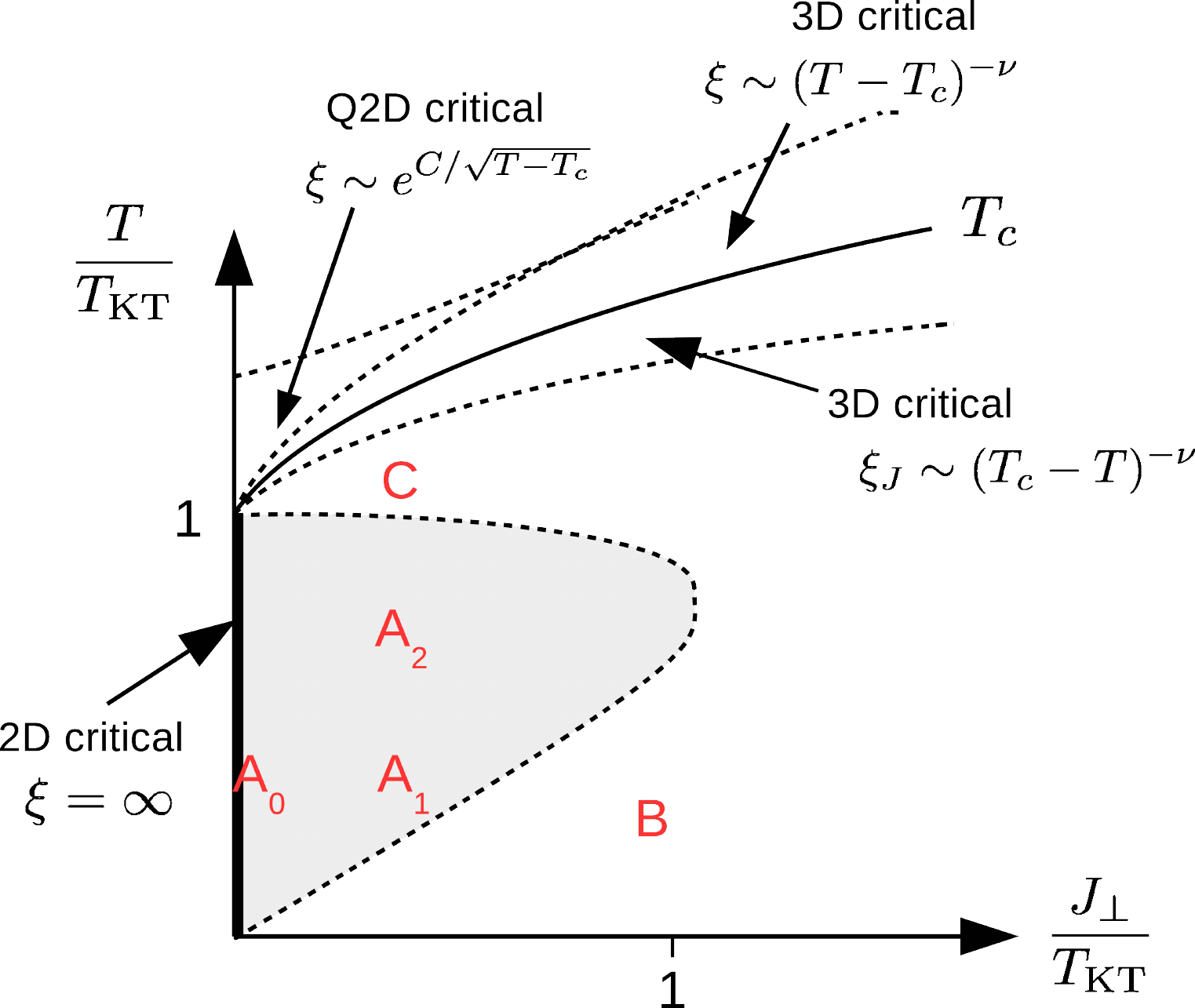}}
\caption{Schematic phase diagram of a layered 3D system with an O(2)-symmetric order parameter as a function of temperature and interplane coupling $\Jperp$. The behavior of the correlation length $\xi$ in the parallel direction near the transition has been studied in previous works and is not discussed in the manuscript.
Our main focus is the existence of a wide region (region A, shown in gray) in the low-temperature phase where signatures of KT physics are observable. Region A is separated from region B, which behaves as a standard anisotropic 3D phase, by a crossover line $T\simeq\Jperp$ where $\Jperp$ is the interplane coupling.} 
\label{fig_phasedia} 
\end{figure}

The main results of our manuscript are summarized in Fig.~\ref{fig_phasedia}. In the low-temperature ordered phase, we can identify three regions, one of which (region A) exhibiting clear signatures of KT physics. This is explained by quasi-2D scaling equations for length scales smaller than the Josephson length $l_J$ defined by the interplane coupling $\Jperp$ 
and the existence of quasi-2D vortex-vortex bound pairs with size $l_v$ much smaller than $l_J$. At low temperatures ($T\lesssim\Tkt$), for length scales between $l_v$ and $l_J$ the RG flow is similar to the critical RG flow which characterizes the low-temperature KT phase dominated by spinwave excitations with a renormalized stiffness. This yields an unusual temperature dependence of the order parameter which varies as a power law of $\Jperp$ with an exponent determined by the anomalous dimension $\eta(T)=1/2\pi\rho_s(T)$ of the strictly 2D KT phase. While this temperature dependence has been known for a long time,\cite{Berezinskii73,Pokrovskii74,Hikami80} we quantitatively determine the region of the phase diagram where it is valid.  
At lower temperatures or for larger values of the interplane coupling (region B), the anisotropy does not play an important role and the 
system behaves essentially as a 3D isotropic system with an order parameter varying quadratically as a function of temperature. At higher temperatures (region C), vortex loops become three dimensional with a size larger than the Josephson length so that signatures of the low-temperature KT phase disappear. 

The outline of the paper is as follows. In Sec.~\ref{sec_modelnprg} we introduce the quasi-2D quantum O(2) model and briefly describe the NPRG approach. In Sec.~\ref{sec_transition}, we discuss the transition temperature and the critical behavior. In Sec.~\ref{sec_lowT} we show how the RG flow allows us to distinguish three regions in the low-temperature phase (regions A, B and C in Fig.~\ref{fig_phasedia}). By computing the temperature dependence of the order parameter, we estimate the boundaries of region A where the order parameter varies as a power law of $\Jperp$ with an exponent which depends on the anomalous dimension of the strictly 2D KT phase.\cite{Berezinskii73,Pokrovskii74,Hikami80}  Region A is separated from region B, which behaves as a standard anisotropic 3D phase (with a quadratic temperature dependence of the order parameter), by a crossover line $T\simeq\Jperp$ where $\Jperp$ is the interplane coupling.

\section{Model and NPRG approach} 
\label{sec_modelnprg}

\subsection{Model}

We consider a quasi-2D quantum O(2) model, describing weakly coupled planes, defined by the action 
\begin{align}
S ={}& \inttau \int d^2r \biggl\{ \sum_i \biggl[ \half(\nablabf\varphibf_i)^2 + \frac{1}{2c^2} (\dtau\varphibf_i)^2 \nonumber \\ & + \frac{r_0}{2} \varphibf_i^2  + \frac{u_0}{4!} {(\varphibf_i^2)}^2 \biggr] 
- \frac{J_\perp^2}{c^2} \sum_{\mean{i,j}} \varphibf_i\cdot\varphibf_j \biggr\} ,
\label{Smicro}
\end{align} 
where $\varphibf_i(\r,\tau)$ is a two-component real field, $\r$ a 2D coordinate, and the indices $i,j$ refer to the planes. 
$\tau\in [0,\beta]$ is an imaginary time, $\beta=1/T$, and we set $\hbar$, $k_B$ as well as the interplane distance equal to unity. The coupling constants $r_0$ and $u_0$ are temperature independent and $c$ is the (bare) velocity of the $\varphibf$ field. $\Jperp$, which has units of energy, denotes the strength of the coupling between nearest-neighbor planes. The model is regularized by a ultraviolet momentum cutoff $\Lamb$.

Equation~(\ref{Smicro}) can be seen as a quantum (relativistic) generalization of the Lawrence-Doniach model.\cite{Lawrence70} At zero temperature there is a quantum phase transition between a disordered phase ($r_0>r_{0c}$) and an ordered phase ($r_0<r_{0c}$) where the O(2) symmetry is spontaneously broken. We will only consider the case $r_0<r_{0c}$ where the ground state is ordered and ignore the immediate vicinity of $r_{0c}$ where the QCP may play a role. 

From Eq.~(\ref{Smicro}) we can define two characteristic length scales. The first one is the thermal length $l_T=c/2\pi T$ separating a quantum regime at small distance from a classical regime at long distance. The second one is the Josephson length scale $l_J=c/\Jperp$ separating a 2D regime at small distance from a 3D regime at long distance.\footnote{Strictly speaking, the Josephson length is modified by the anomalous dimension $\eta$ of the 2D KT phase. Using $[\Jperp]=1-\eta/2$ (see Sec.~\ref{subsec_op}), one finds $l_J=\Lamb^{-1}(c\Lamb/\Jperp)^{2/(2-\eta)}$. Since $\eta\leq 1/4$, one can approximate $l_J\simeq c/\Jperp$.}
Quantum fluctuations are crucial at low temperatures when $l_T$ is larger than $l_J$ but do not play an important role at sufficiently high temperatures when $l_T\ll l_J$; the quasi-2D quantum O(2) model is then equivalent to the Lawrence-Doniach model and qualitatively describes the physics of the layered 3D XY model. Note however that $c$ and $\Jperp$ are related to the spin-spin couplings $J^{\rm XY}_\para$ and $J^{\rm XY}_\perp$ of the 3D anisotropic XY model by $c/\Jperp=(J^{\rm XY}_\para/J^{\rm XY}_\perp)^{1/2}$.\footnote{This is easily seen by comparing the spinwave propagators of the quasi-2D quantum O(2) model and layered 3D XY model.}

\subsection{NPRG approach}

The strategy of the NPRG is to build a family of theories indexed by a momentum scale $k$ such that fluctuations are smoothly taken into account as $k$ is lowered from the microscopic scale $\Lamb$ down to zero.\cite{Berges02,Delamotte12,Kopietz_book} This is achieved by adding to the action~(\ref{Smicro}) the infrared regulator
\begin{equation}
\Delta S_k[\varphibf] = \half \sum_p \varphibf(-p) R_k(p) \varphibf(p) , 
\end{equation}
where we use the notation $p=(\p,i\wn)=(\p_\para,p_\perp,i\wn)$ with $\wn=2n\pi T$ ($n$ integer) a Matsubara frequency. $\p_\para$ denotes the 2D component of $\p$ parallel to the planes while $p_\perp$ is the perpendicular component. The regulator function $R_k$ will be specified below. The partition function 
\begin{equation}
\calZ_k[\J] = \int \calD[\varphibf] \, e^{-S[\varphibf]-\Delta S_k[\varphibf] + \inttau \int d^2r \sum_i \J_i\cdot\varphibf_i } ,
\end{equation}
defined here in the presence of an external source $\J$, is $k$ dependent. The scale-dependent effective action 
\begin{equation}
\Gamma_k[\phibf] = - \ln\calZ_k[\J] + \inttau \int d^2r \sum_i \J_i\cdot\phibf_i - \Delta S_k[\phibf] 
\end{equation} 
is defined as a modified Legendre transform of $-\ln \calZ_k[\J]$ which includes the subtraction of $\Delta S_k[\phibf]$. Here $\phibf=\mean{\varphibf}$ is the order parameter (in the presence of the external source). 

Assuming fluctuations to be frozen by the $\Delta S_k$ term at the microscopic scale $\Lamb$, the initial condition of the flow is then simply given by $\Gamma_\Lamb[\phibf]=S[\phibf]$, which reproduces mean-field theory. The effective action of the original model is given by $\Gamma_{k=0}[\phibf]$ provided that $R_{k=0}$ vanishes. For a generic value of $k$, the regulator $R_k(p)$ suppresses low-energy fluctuations but leaves high-energy fluctuations essentially unaffected (see below).\cite{Berges02,Delamotte12,Kopietz_book}

The variation of the effective action with $k$ is given by Wetterich's equation,\cite{Wetterich93} 
\begin{equation}
\dt \Gamma_k[\phibf] = \half \Tr\bigl\{ \dt R_k (\Gamma^{(2)}_k[\phibf] + R_k )^{-1} \bigr\} , 
\label{weteq} 
\end{equation}
where $t=\ln(k/\Lamb)$ is a (negative) RG time and $\Gamma_k^{(2)}$ denotes the second-order functional derivative of $\Gamma_k$. In Fourier space the trace involves a sum over momenta and frequencies as well as the internal index of the two-component field $\phibf$. 

To solve the RG equation~(\ref{weteq}) we use a derivative expansion of $\Gamma_k[\phibf]$ (which we refer to as LPA$'$ where LPA stands for local potential approximation),  
\begin{multline}
\Gamma_k[\phibf] = \inttau \int d^2r  \biggl\{ \sum_i \biggl[ \frac{Z_{\para,k}}{2}(\nablabf\phibf_i)^2 + \frac{Z_{\tau,k}}{2} (\dtau\phibf_i)^2 \\ + U_k(\rho_i) + Z_{\perp,k} \phibf_i\cdot \phibf_i\biggr] - Z_{\perp,k} \sum_{\mean{i,j}} \phibf_i\cdot\phibf_j \biggr\} ,
\label{gammadef} 
\end{multline}  
which includes only nearest-neighbor-plane coupling. For symmetry reasons, the effective potential $U_k$ is a function of the O(2) invariant $\rho=\phibf^2/2$. We expand $U_k(\rho)$ about its minimum at $\rho=\rho_{0,k}$,
\begin{equation}
U_k(\rho) = U_{0,k} + \delta_k(\rho-\rho_{0,k}) + \frac{\lamb_k}{2} (\rho-\rho_{0,k})^2 .
\label{Udef} 
\end{equation}
In the ordered phase, where the O(2) symmetry is spontaneously broken, $\rho_{0,k}$ is nonzero and $\delta_k=U'_k(\rho_{0,k})=0$. In the disordered phase $\rho_{0,k}$ vanishes and $\delta_k=U'_k(0)\neq 0$. From~(\ref{gammadef}) we can obtain the longitudinal and transverse parts of the two-point vertex, 
\begin{equation}
\begin{split} 
\Gamma^{(2)}_{k,\rm T}(p) &= Z_{\para,k}\p_\para^2 + Z_{\tau,k} \wn^2 + Z_{\perp,k} \eps_\perp(p_\perp) + \delta_k , \\ 
\Gamma^{(2)}_{k,\rm L}(p) &= \Gamma^{(2)}_{\rm T,k}(p) + 2 \lamb_k \rho_{0,k} ,  
\end{split}
\end{equation}
where $\eps_\perp(p_\perp)=2(1-\cos p_\perp)$, and in turn the transverse and longitudinal (wrt to the direction of the order parameter) parts of the propagator: $G_{k,\alpha}(p)=1/\Gamma^{(2)}_{k,\alpha}(p)$ ($\alpha={\rm T,L}$). The (running) longitudinal and transverse anomalous dimensions are defined by 
\begin{equation}
\eta_{\para,k}=-\dt \ln Z_{\para,k} , \quad \eta_{\perp,k}=-\dt \ln Z_{\perp,k} .
\label{etadef} 
\end{equation}

In the low-energy limit, using $\eps_\perp(p_\perp)\simeq p^2_\perp$ for $|p_\perp|\ll 1$, the spectrum of transverse fluctuations is therefore given by 
\begin{equation}
\w^2_\p = c^2_{\para,k} \p_\para^2 + c^2_{\perp,k} p^2_\perp + \frac{\delta_k}{Z_{\tau,k}} 
\end{equation}
with anisotropic velocities
\begin{equation}
c_{\para,k} = \sqrt{\frac{Z_{\para,k}}{Z_{\tau,k}}} , \quad c_{\perp,k} = \sqrt{\frac{Z_{\perp,k}}{Z_{\tau,k}}} .
\end{equation}
In agreement with Goldstone's theorem, the transverse (spinwave) spectrum is gapless in the ordered phase ($\delta_k=0$).

We are now in a position to define the regulator function:  
\begin{equation}
\begin{split} 
R_k(p) &= Z_{\para,k} k^2 Y r(Y) , \quad  r(Y) = \frac{\alpha}{e^Y-1} , \\ 
Y &= \frac{1}{k^2} \biggl[ \p_\para^2 + \frac{\wn^2}{c^2_{\para,k}} + \frac{c^2_{\perp,k}}{c^2_{\para,k}} \eps_\perp(p_\perp) \biggr] , 
\end{split} 
\end{equation}
with an arbitrary parameter $\alpha$ of order unity (we choose $\alpha=1$). The derivation of the flow equations is standard.\cite{[{See, e.g., }]Rancon13a} The latter are given in the Appendix.

\section{Phase transition and critical behavior} 
\label{sec_transition}

\begin{figure}
\centerline{\includegraphics[width=8cm]{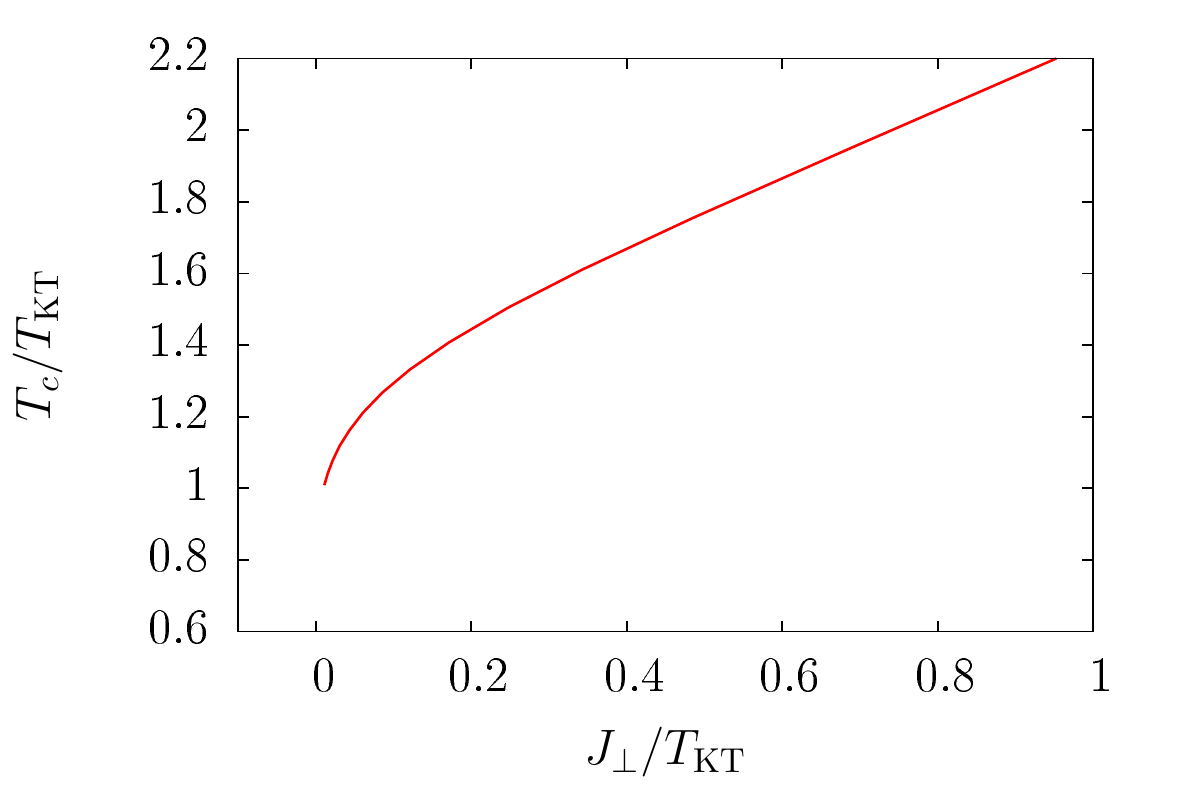}}
\caption{Transition temperature vs $\Jperp$.} 
\label{fig_Tc} 
\end{figure}

\begin{figure}
\centerline{\includegraphics[width=8cm]{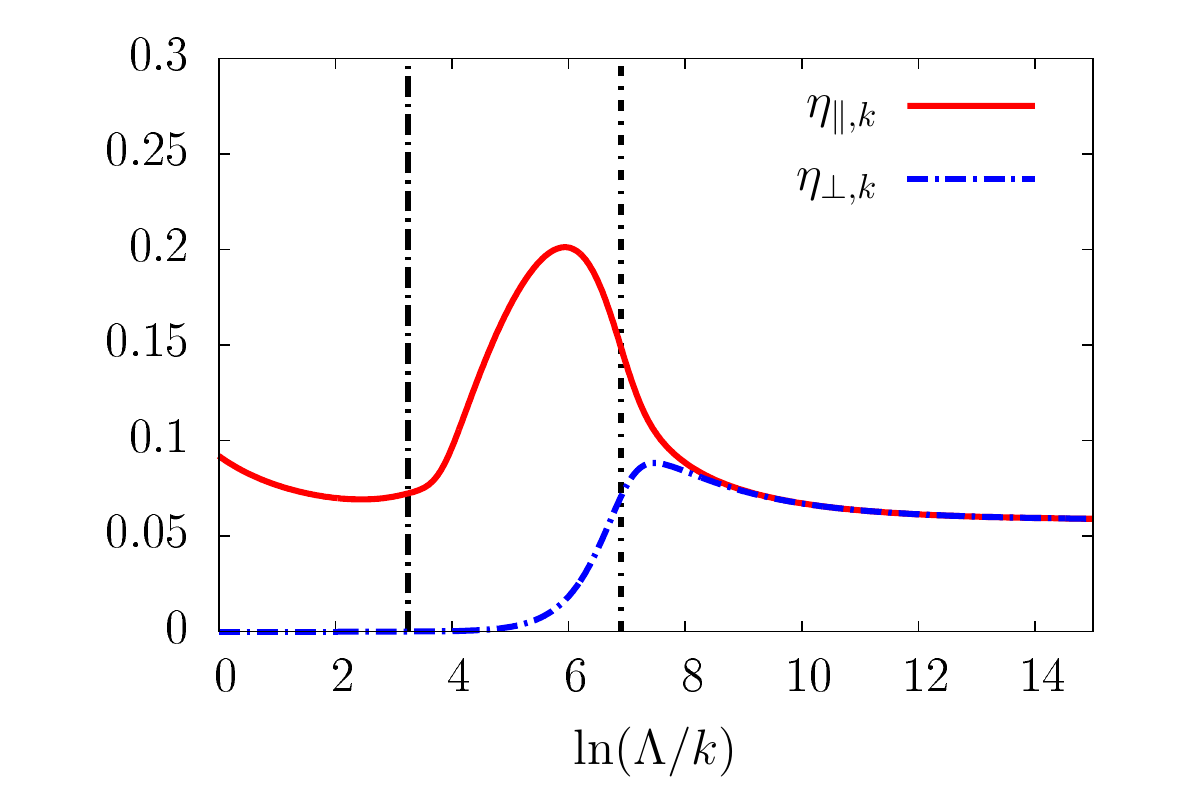}}
\caption{Anomalous dimensions $\eta_{\para,k}$ and $\eta_{\perp,k}$ vs $\ln(\Lamb/k)$ at the transition temperature $T_c\simeq 1.5\Tkt$ for $\Jperp=\Tkt/4$. The two vertical lines correspond to the thermal scale $k_T=1/l_T$ and Josephson scale $k_J=1/l_J$.}
\label{fig_eta_Tc} 
\end{figure}

The transition temperature obtained from the NPRG is shown in Fig.~\ref{fig_Tc}. For $\Jperp\to 0$, $T_c$ tends to a nonzero value that we identify with the KT transition temperature $\Tkt$ of the 2D model. $T_c(\Jperp\to 0^+)$ is indeed compatible with the KT transition temperature that can be inferred from the 2D RG flow even though the NPRG predicts the 2D system to be always disordered (see the discussion in the introduction).\footnote{Using the standard method to estimate $\Tkt$ from the 2D NPRG flow in the LPA$'$ (see, e.g., Refs.~\onlinecite{Rancon12b,Rancon13b}), we find $\Tkt/c\Lamb\simeq 0.004$ in agreement with $T_c(\Jperp\to 0^+)$.} For $\Jperp\to 0^+$, we expect 
\begin{equation}
T_c - \Tkt \sim \frac{c}{\left(\ln \frac{c}{\Jperp}\right)^2 } .
\label{Tc}
\end{equation} 
This result is obtained from the criterion $\xi_{\rm 2D}\sim l_J$ where $\xi_{\rm 2D}\sim \exp(\const/\sqrt{T-\Tkt})$ is the correlation length  of the 2D model.\cite{Hikami80} Our approach, which does not reproduce the scaling of the 2D correlation length $\xi_{\rm 2D}$ with $T-\Tkt$, does not yield Eq.~(\ref{Tc}) even though we do obtain an upward shift of $T_c$ for $\Jperp>0$ as expected. 

On the other hand we clearly see a 3D/2D crossover in the RG flow (Fig.~\ref{fig_eta_Tc}). For $kl_J\gtrsim 1$, the RG flow is essentially 2D as can be seen from the vanishing of the transverse anomalous dimension: $\eta_{\perp,k}\simeq 0$, i.e. $\dk Z_{\perp,k}\simeq 0$. For $kl_J\sim 1$, there is a crossover to a critical 3D regime where both $\eta_{\para,k}$ and $\eta_{\perp,k}$ are nonzero and close to the known value $\eta^*\simeq 0.0584$ at the 3D Wilson-Fisher fixed point in the LPA$'$ with a truncated effective potential [Eq.~(\ref{Udef})].

\section{Low-temperature phase} 
\label{sec_lowT} 

In this section we show that the low-temperature phase bears clear signatures of KT physics in a wide region of the phase diagram. We first discuss how Q2D KT physics manifests itself in the RG flow before considering the temperature dependence of the order parameter.

\subsection{RG flows and characteristic scales} 

\begin{figure} 
\centerline{\includegraphics[width=8.5cm]{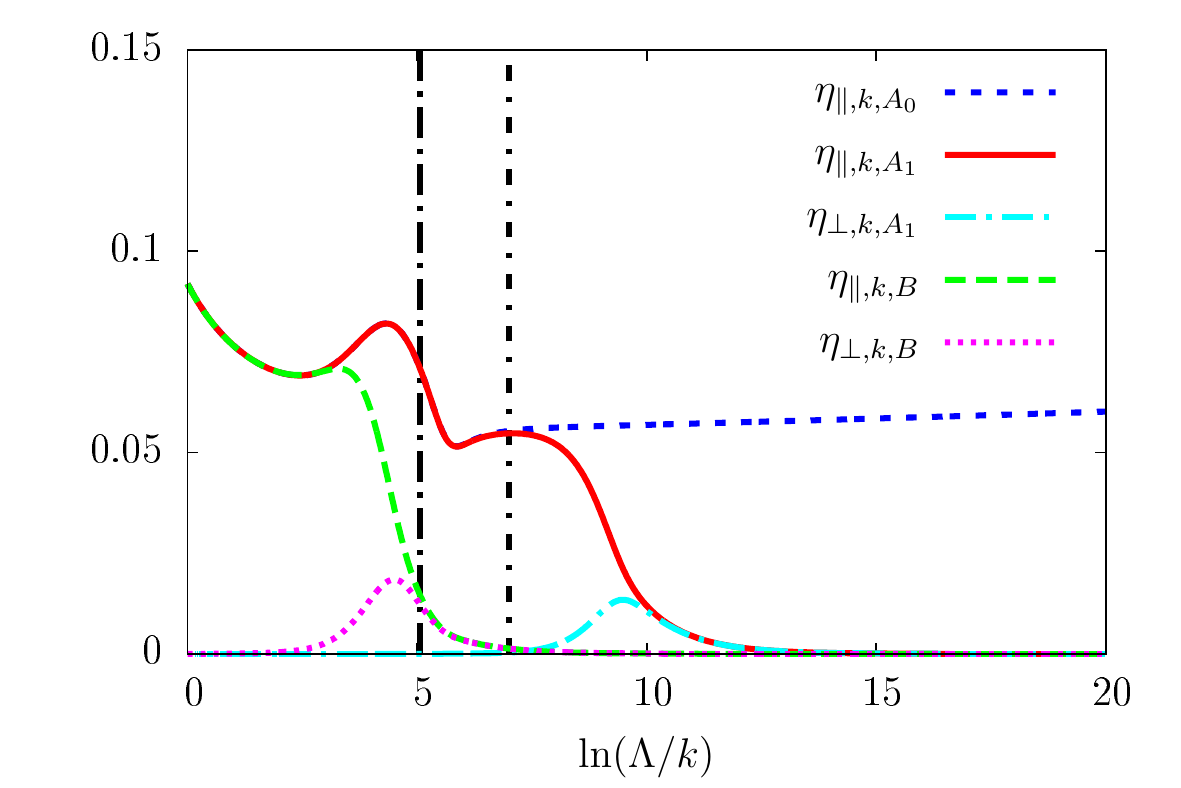}}
\caption{Anomalous dimensions $\eta_{\para,k}$ and $\eta_{\perp,k}$ vs $\ln(\Lamb/k)$ for $T=\Tkt/4$ and $\Jperp=0,\Tkt/40,2.5\Tkt$ corresponding to the  points $A_0,A_1,B$ loosely shown in Fig.~\ref{fig_phasedia}. The vertical lines correspond to the thermal scale $k_T$ ($\ln(\Lamb l_T)\simeq 5.1$) and to the scale $k_v$ associated with the size $l_v$ of the largest vortex-vortex bound pairs. The maximum of $\eta_{\perp,k}$ corresponds to the Josephson scale $l_J$ ($\ln(\Lambda l_J)\simeq 9.2$ for $A_1$ and $\ln(\Lambda l_J)\simeq 4.6$ for $B$).}   
\label{fig_eta_AB} 
\centerline{\includegraphics[width=8.5cm]{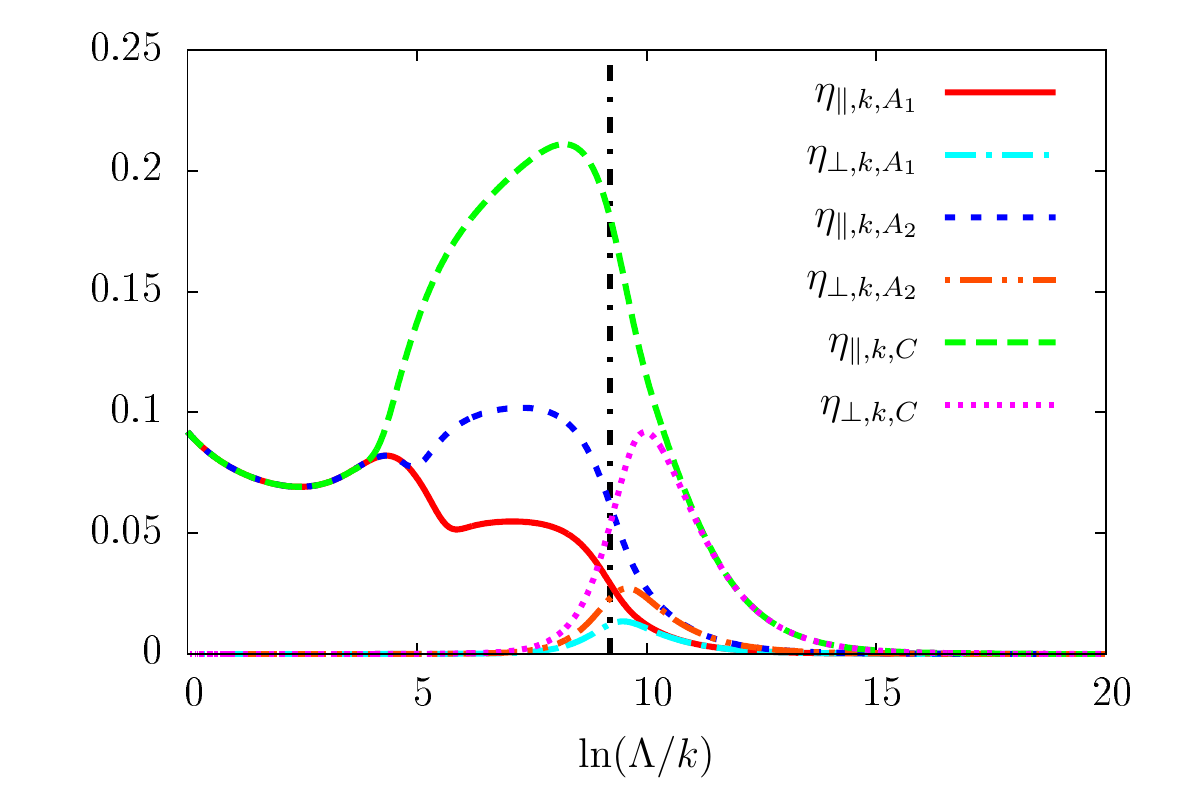}}
\caption{Same as Fig.~\ref{fig_eta_AB} but for $\Jperp=\Tkt/40$ and $T=\Tkt/4,\Tkt/2,1.05\Tkt$ (points $A_1,A_2,C$ loosely shown in  Fig.~\ref{fig_phasedia}). The vertical line corresponds to the Josephson scale ($\ln(\Lambda l_J)\simeq 9.2$).}
\label{fig_eta_AC} 
\end{figure}

Let us consider the RG flow for $T<\Tkt\leq T_c$. Figures~\ref{fig_eta_AB} and \ref{fig_eta_AC} show the behavior of the anomalous dimensions $\eta_{\para,k}$ and $\eta_{\perp,k}$ as we go from point $A_0$ ($\Jperp=0$ and $T=\Tkt/4$) of the phase diagram (Fig.~\ref{fig_phasedia}) to point $A_1$ ($\Jperp=\Tkt/40$ and $T=\Tkt/4$) and then to point B ($\Jperp=2.5\Tkt$ and $T=\Tkt/4$), or points $A_2$ ($\Jperp=\Tkt/40$ and $T=\Tkt/2$) and $C$ ($\Jperp=\Tkt/40$ and $T=1.05\Tkt<T_c$, $T_c\simeq 1.1.\Tkt$). 

In the 2D limit $\Jperp=0$ (point $A_0$) $\eta_{\perp,k}=0$ and after a transient regime $\eta_{\para,k}$ shows a (quasi) plateau in agreement with the expectation that the system is critical for $T\leq \Tkt$ with a nonzero anomalous dimension $\eta\equiv\eta_{\para,k}^{\rm plateau}$ and a nonzero stiffness $\rho_s=1/2\pi\eta$ (Fig~\ref{fig_eta_AB}). It is convenient to introduce a characteristic length scale $l_v\equiv l_v(T)$ associated with the size of the largest vortex-vortex bound pairs. While the flow in the range $k\gtrsim l^{-1}_T$ is a quantum $(2+1)$D flow, we can identify a window $l^{-1}_v\lesssim k\lesssim l^{-1}_T$ with a 2D (thermal) flow where both spinwave and vortex excitations are important.\footnote{We see in Fig.~\ref{fig_eta_AB} that the effect of temperature is actually visible for $k$ slightly smaller than $k_T=c/2\pi T$. 
This means that the ``true'' thermal scale $k_T$ is very close to $k_v$ (defined by the beginning of the plateau in $\eta_{\para,k}$) so that the window $l^{-1}_v\lesssim k\lesssim l^{-1}_T$ is hard to see for $T=\Tkt/4$. It however widens with increasing temperature.}
Once $k$ becomes smaller than $l^{-1}_v$, the flow is entirely determined by the spinwave excitations and the anomalous dimension $\eta_{\para,k}\simeq \eta$ takes a constant value. Note that we are not able {\it a priori} to estimate $l_v$, its value is simply inferred from the beginning of the plateau of $\eta_{\para,k}$. 
In the following we refer to the flow in the range $k\lesssim l_v^{-1}$ as the critical KT flow.   

We are now in a position to investigate the effect of a nonzero interplane coupling $\Jperp$. Let us first move from point $A_0$ to $A_1$ in the phase diagram (Fig.~\ref{fig_phasedia}). $A_0$ and $A_1$ correspond to the same temperature $T=\Tkt/4$ but $\Jperp=\Tkt/40$ is nonzero for $A_1$. We see that the RG flows are identical for $A_0$ and $A_1$ as long as $k\gtrsim l^{-1}_J$, where $l_J$ is the Josephson length for $\Jperp=\Tkt/40$ (Fig.~\ref{fig_eta_AB}).
Vortex-vortex bound pairs with size $<l_v$ must now be interpreted as Q2D rectangular vortex loops, i.e. long rectangular loops perpendicular to the planes and cutting a single plane.\cite{Hikami80,Shenoy89,Shenoy95} When $k\sim l_J^{-1}$, $\eta_{\para,k}$ is strongly suppressed and $\eta_{\perp,k}$ becomes nonzero thus signaling a crossover to a 3D regime. For $k\ll l^{-1}_J$, both $\eta_{\para,k}$ and $\eta_{\perp,k}$ vanish while the stiffnesses $Z_{\para,k},Z_{\perp,k}$ and the order parameter $\rho_{0,k}$ (not shown) take a nonzero constant value, as expected in a 3D ordered phase. 

From point $A_1$, let us now move to point $B$ by increasing again $\Jperp$ at fixed temperature $T=\Tkt/4$. Because of the large value $\Jperp=2.5\Tkt$ at point $B$, the Josephson length is now smaller than the characteristic vortex length $l_v$. As a result, the critical KT flow which was observed at point $A_1$ for $l_J^{-1}\lesssim k\lesssim l^{-1}_v$ has disappeared (Fig.~\ref{fig_eta_AB}). We therefore do not expect any KT signatures in physical quantities.

Let us finally consider points $A_2$ and $C$ obtained from $A_1$ by increasing temperature at fixed $\Jperp=\Tkt/40$ (Fig.~\ref{fig_phasedia}). At point $A_2$ ($T=\Tkt/2$), we can still observe a critical KT flow characterized by a plateau in the anomalous dimension $\eta_{\para,k}$ (Fig.~\ref{fig_eta_AC}). This plateau corresponds to a larger value than for $A_1$, which reflects the increase with temperature of the anomalous dimension in the 2D phase. On the other hand, at point $C$ the temperature $T=1.05\Tkt$ is larger than $\Tkt$ (but nevertheless smaller than $T_c\simeq 1.1\Tkt$) and the Q2D critical KT flow at intermediate regime of $k$ has disappears: there is no plateau in $\eta_{\para,k}$ anymore and the anomalous dimension rises above 0.2 thus getting close of the anomalous dimension $\eta=0.25$ at the KT transition in the 2D system. This observation agrees with the expectation that, when increasing temperature, the maximum size $l_v$ of Q2D rectangular vortex loops increases and eventually becomes of the order of $l_J$, thus suppressing the Q2D KT physics. Since $l_v$ diverges at $\Jperp=0$ when $T$ becomes of order $\Tkt$, we expect the temperature at which $l_v\sim l_J$ to be of the order of $\Tkt$ for small $\Jperp$ (as confirmed by the numerical results). 

These results are sufficient to delimit a region of the phase diagram where $l_v<l_J$ so that a Q2D critical KT flow at intermediate values of $k$ exists. On the one hand, as illustrated by Fig.~\ref{fig_eta_AB}, at low temperatures where $l_v\sim l_T$ the condition $l_v<l_J$ implies $\Jperp\lesssim T$. At higher temperatures, as illustrated by Fig.~\ref{fig_eta_AC}, the same condition implies $T\lesssim\Tkt$. Only when these two conditions, $T\lesssim\Tkt$ and $\Jperp\lesssim T$, are fulfilled, does a Q2D critical KT flow exist at intermediate values of $k$. The crossover line $T\sim\Tkt$ between regions A and C has also been discussed by Shenoy {\it et al.}\cite{Shenoy95} We shall see in the following section that whenever a Q2D critical KT flow exists, the order parameter shows a power-law temperature dependence of the order parameter with an exponent which depends on the anomalous dimension $\eta\equiv\eta(T)$ of the 2D KT phase.

\subsection{Order parameter} 
\label{subsec_op} 

\subsubsection{Classical regime $\Jperp\lesssim T\lesssim \Tkt$} 

Since the order parameter $\sqrt{\rho_0}$ is zero for $\Jperp=0$, it must vanish as a power law when $\Jperp\to 0$ and $T<\Tkt$.
This power law can be simply obtained by scaling analysis. In the low-temperature critical KT phase, the field has scaling dimension $[\varphibf]=\half(d-2+\eta)=\eta/2$ where $d=2$ is here the space dimension of the system when $\Jperp=0$ and $\eta\equiv\eta(T)$ the anomalous dimension. Since $[\Jperp]=1-\eta/2$, we deduce\cite{Berezinskii73,Pokrovskii74,Hikami80}
\begin{equation}
\sqrt{\rho_0} = \sqrt{\calA(T)} \left( \frac{\Jperp}{T} \right)^\frac{\eta}{2-\eta} \quad \mbox{for} \quad \frac{\Jperp}{T}\ll 1  .
\label{rho0}
\end{equation}
The origin of the temperature-dependent energy scale $\calA(T)$ is twofold. First it takes into account the fact that classical fluctuations giving rise to the power-law dependence in $\Jperp$ are limited to length scales beyond the thermal length $l_T=c/2\pi T$. Second it takes into account quantum fluctuations occurring at small length scales ($<l_T$).
$\Jperp$ shows up in the combination $\Jperp/T$ since the interplane coupling of the effective classical model describing the low-energy physics is $\Jperp/T$.\footnote{At length scales larger than the thermal length $l_T$, only the vanishing Matsubara frequency $\w_{n=0}$ contributes to the action. The integration over the imaginary time then simply yields a factor $\beta$, and $\beta\Jperp=\Jperp/T$ becomes the interplane coupling of the classical quasi-2D O(2) model.}
We shall see that Eq.~(\ref{rho0}) holds whenever the system exhibits a Q2D critical KT flow and is therefore satisfied in a wide region of the phase diagram. Equation~(\ref{rho0}) is valid only for $T\lesssim\Tkt$. In the immediate vicinity of the transition temperature $T_c$, $\sqrt{\rho_0}\sim (T_c-T)^\beta$ satisfies the usual scaling law with $\beta$ the critical exponent associated with the 3D XY universality class. This result is confirmed by our NPRG calculation with $\beta\simeq 0.324$ in agreement with the known value in the LPA$'$ with a truncated effective potential. 
 
\begin{figure}
\centerline{\includegraphics[width=8cm]{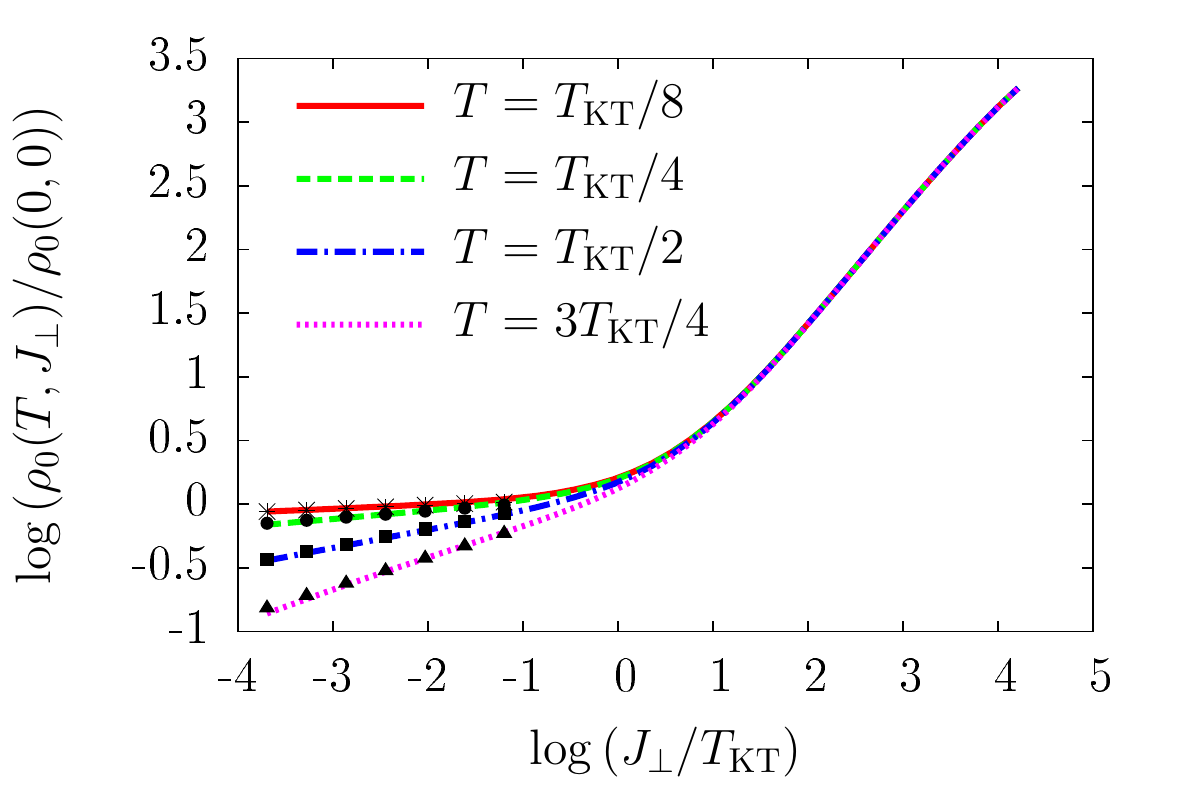}}
\caption{Order parameter $\rho_0$ vs $\Jperp$ for various temperatures. The (black) symbols correspond to the power-law dependence~(\ref{rho0}).} 
\label{fig_rho0}
\end{figure}

Figure~\ref{fig_rho0} shows the $\Jperp$-dependence of the order parameter $\sqrt{\rho_0}$ for various values of $T$ below $\Tkt$. The RG result is compared with Eq.~(\ref{rho0}) where the anomalous dimension $\eta\equiv\eta(T)$ is obtained from the flow at $\Jperp=0$. At low temperatures $T\ll\Tkt$ and for $\Jperp=0$, $\eta_{\para,k}$ shows a well-defined plateau and it is possible to determine $\eta$ with good precision. At higher temperatures, the plateau is not so well defined\footnote{This is due to the fact that our NPRG approach does not yield a line of fixed points {\it stricto sensu} but rather a line of quasi-fixed points.} (in Fig.~\ref{fig_eta_AB} we see that already for $T=\Tkt/4$ the ``plateau'' of $\eta_{\para,k}$ shows a nonzero slope) and only a rough estimate of $\eta$ can be obtained. Nevertheless, our numerical results for $\rho_0$ show a very good agreement with the expected power-law behavior~(\ref{rho0}) when $\Jperp\lesssim T$. In Sec.~\ref{subsec_crossover} we shall use the results shown in Fig.~\ref{fig_rho0} to define more precisely the crossover line between regions A and B in Fig.~\ref{fig_phasedia}.

\subsubsection{Quantum regime $T\lesssim\Jperp$} 

In the low-temperature regime the system behaves essentially as a 3D system and we expect the one-loop approximation of the effective action $\Gamma[\phibf]$ (equivalent to the spinwave approximation of the layered 3D  XY model) to give an accurate estimate of the order parameter. At the mean-field level, $\Gamma[\phibf]=S[\phibf]$, so that the mean-field order parameter is defined by 
\begin{equation}
\rho_0^{\rm MF} = - \frac{3}{u_0} \left( r_0 -  2 \frac{\Jperp^2}{c^2} \right) . 
\end{equation}
To one-loop order, 
\begin{align}
\rho_0 ={}& \rho_0^{\rm MF} - \half \int_q [ 3 G_{\rm L}(q) + G_{\rm T}(q) ] , 
\end{align} 
where 
\begin{equation}
\begin{split} 
G_{\rm T}(q) &= \Bigl[ \q_\para^2 + \frac{\wn^2}{c^2} +  \frac{\Jperp^2}{c^2} \eps_\perp(q_\perp)  \Bigr]^{-1} , \\
G_{\rm L}(q) &= \Bigl[ \q_\para^2 + \frac{\wn^2}{c^2} +  \frac{\Jperp^2}{c^2} \eps_\perp(q_\perp) + \frac{2}{3} u_0 \rho_0^{\rm MF} \Bigr]^{-1} 
\end{split}
\end{equation}
are the mean-field propagators and we use the notation 
\begin{equation}
\int_q = T \sum_{\wn} \int \frac{d^2 q_\para}{(2\pi)^2} \int_{-\pi}^{\pi} \frac{d\qperp}{2\pi} .
\end{equation}
When $T\ll\Jperp$, the temperature dependence of the order parameter is dominated by the (gapless) transverse fluctuations, 
\begin{equation}
\rho_0(T)-\rho_0(0) \simeq - \half \biggl[ \int_q G_{\rm T}(q) - \int_q G_{\rm T}(q) \Bigl|_{T=0} \biggr] ,
\end{equation}
where we can approximate $\eps_\perp(q_\perp)\simeq q_\perp^2$ and extend the $q_\perp$ integral from $-\infty$ to $\infty$. A straightforward calculation gives 
\begin{equation}
\rho_0(T)-\rho_0(0) \simeq - \frac{T^2}{24\Jperp} . 
\label{rhosw}
\end{equation}
Equation~(\ref{rhosw}) ignores the renormalization of $\Jperp$ by quantum fluctuations. To improve on Eq.~(\ref{rhosw}) we introduce the effective (renormalized) value $\Jperp^{\rm eff}=(c^2_k Z_{\perp,k})^{1/2}|_{k\ll l_J^{-1}}$,\footnote{Recall that $\eta_{\para,k}\simeq 0$ and $\eta_{\perp,k}\simeq 0$ for $k\ll l_J^{-1}$ so that $(c^2_k Z_{\perp,k})^{1/2}$ becomes $k$ independent.}   
which leads to 
\begin{equation}
\rho_0(T)-\rho_0(0) \simeq - \frac{T^2}{24 \Jperp^{\rm eff}} . 
\label{rhosw1}
\end{equation} 

\subsubsection{Temperature dependence of the order parameter}

\begin{figure}
\centerline{\includegraphics[width=8cm]{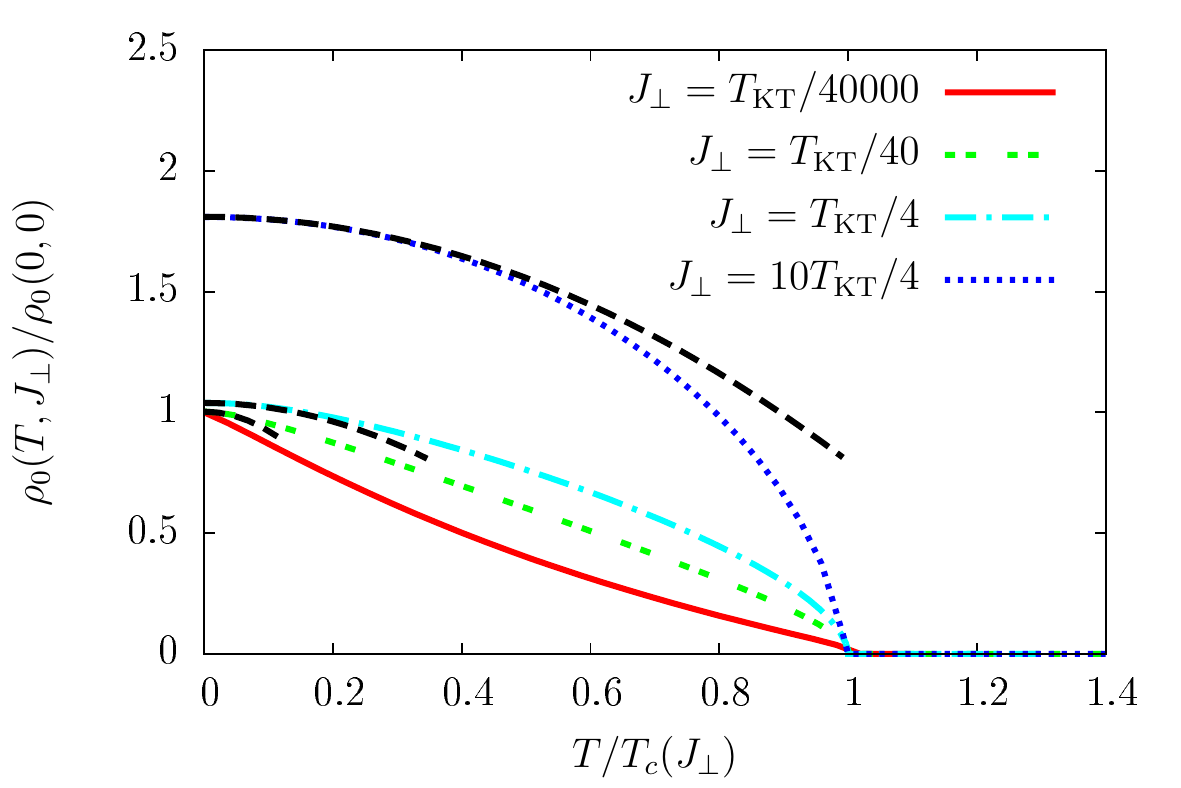}}
\caption{Temperature dependence of the order parameter for various values of $\Jperp$. The (black) dashed lines show the expected low-temperature result~(\ref{rhosw1}).}
\label{fig_rho0T}
\centerline{\includegraphics[width=8cm]{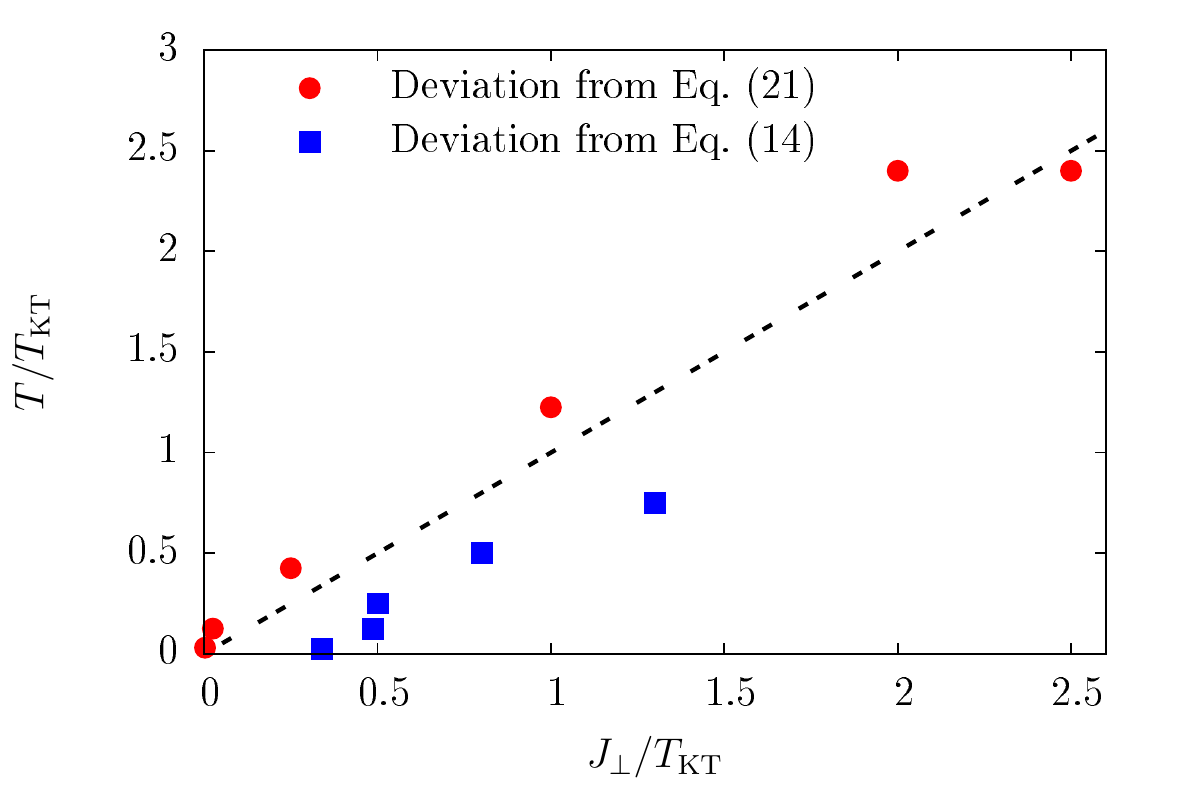}} 
\caption{Crossover line between regions A and B (see Fig.~\ref{fig_phasedia}) obtained by comparing the NPRG result for $\rho_0(\Jperp,T)$ to the small-$\Jperp$ expression~(\ref{rho0}) or the low-temperature limit~(\ref{rhosw1}). The dashed line corresponds to $T=\Jperp$.}  
\label{fig_crossover} 
\end{figure}

Figure~\ref{fig_rho0T} shows the overall temperature dependence of the order parameter $\rho_0$ as obtained from the NPRG approach. For large values of $\Jperp$, e.g. $\Jperp=2.5\Tkt$,\footnote{Note that $J_\perp = 2.5T_{\rm KT}$ corresponds to a small interplane coupling, $J_\perp/J_\para\simeq 0.01$, with an intraplane coupling $J_\para\sim c\Lambda$.}
the curve $\rho_0(T)$ shows a pronounced concavity (as in the mean-field theory) with a sizable low-temperature regime where the quadratic dependence~(\ref{rhosw1}) is observed. For lower values of $\Jperp$, the concavity is less pronounced and the low-temperature regime with a quadratic dependence shrinks. For extremely small values of $\Jperp$,  the curve becomes slightly convex and the low-temperature quadratic regime is not visible any more.

The temperature dependence of the order parameter has recently been studied in a model of strongly anisotropic quantum antiferromagnets.\cite{Furuya16} A pronounced convexity of $\rho_0(T)$ for small $\Jperp$ was found and the low-temperature quadratic variation of the order parameter was not (always) reproduced (see Figs.~3, 10 and 12 in Ref.~\onlinecite{Furuya16}). Without the temperature-dependent energy scale $\calA(T)$, we  would obtain from Eq.~(\ref{rho0}) a pronounced convexity for small $\Jperp$ as in Ref.~\onlinecite{Furuya16}. A possible explanation for the difference between our results and those of Ref.~\onlinecite{Furuya16} is therefore the role of quantum fluctuations, which were not taken into account in the classical approximation used in Ref.~\onlinecite{Furuya16}.

\subsection{Estimate of the size of region A} 
\label{subsec_crossover}

From the results of Fig.~\ref{fig_rho0}, we define the crossover line between regions A and B by a 5\% difference in the order parameter $\rho_0$ between the NPRG results of Fig.~\ref{fig_rho0} and the small $\Jperp$ behavior~(\ref{rho0}). Another estimate of the crossover line can be obtained from a 5\% difference between the result of Fig.~\ref{fig_rho0T} and the low-temperature behavior~(\ref{rhosw1}). The crossover line, shown in Fig.~\ref{fig_crossover}, is roughly defined by $T\simeq\Jperp$.

\subsection{Stiffness} 

\begin{figure}
\centerline{\includegraphics[width=8cm]{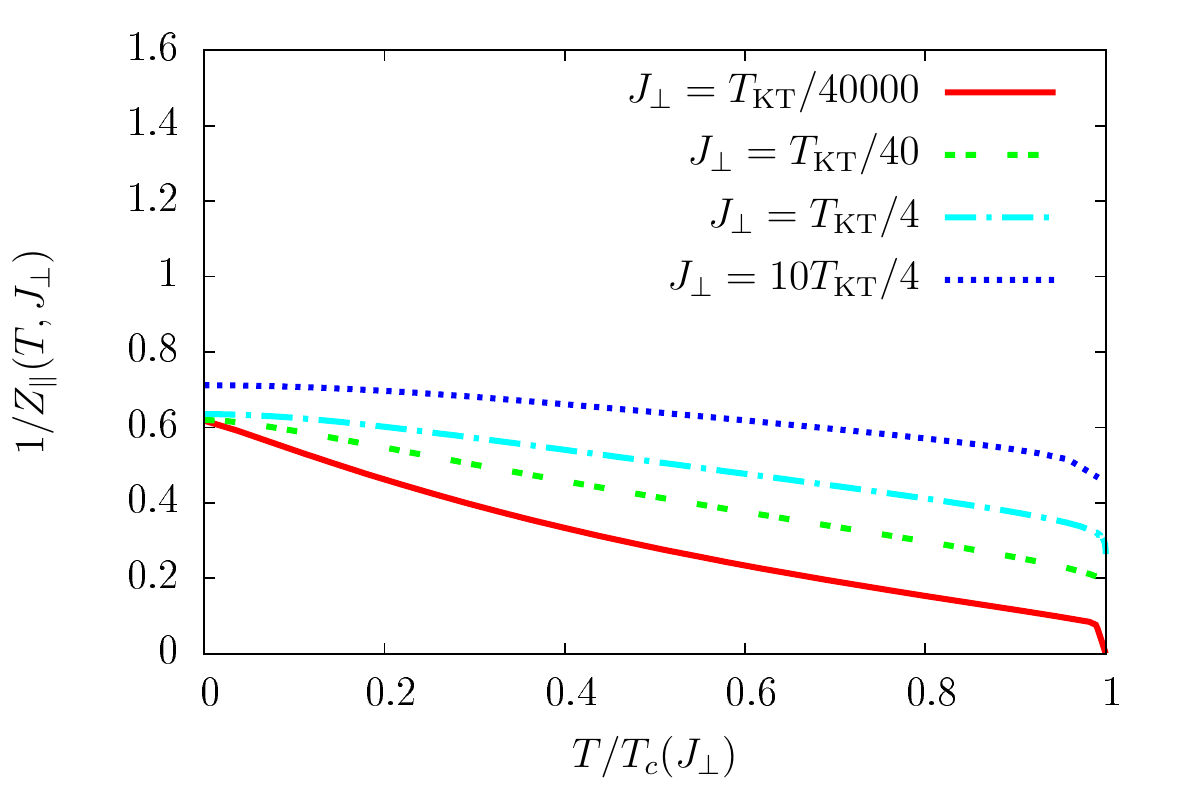}}
\caption{Temperature dependence of the ratio $2\rho_0/\rho_s=1/Z_{\para}(T,\Jperp)$ between the square of the order parameter and the in-plane stiffness for various values of $\Jperp$.}
\label{fig_rho0rhos}
\end{figure}

Figure~\ref{fig_rho0rhos} shows the ratio between the square of the order parameter $2\rho_0$ and the in-plane stiffness $\rho_s=2\rho_0Z_\para$  corresponding to the ratio between the condensate density and the superfluid density in a bosonic superfluid. In a weakly interacting three-dimensional Bose gas, this ratio is close to unity except in the immediate vicinity of the superfluid transition where it vanishes as $(T_c-T)^{\nu\eta}$ where $\nu\simeq 0.6719$ and $\eta\simeq 0.03852$ are critical exponents of the three-dimensional O(2) universality class.\cite{Kos16}
In the 2D limit below $\Tkt$ it is equal to zero since the order parameter vanishes while the stiffness is finite. In the quasi-2D case, $2\rho_0/\rho_s$ is close to unity for small anisotropy but strongly suppressed when $\Jperp\ll\Tkt$.

\section{Conclusion}

In conclusion we have studied the quantum Lawrence-Doniach model using the NPRG. Even though we do not introduce vortices explicitly, our results can be understood by considering, in addition to the Josephson length $l_J$, a characteristic length $l_v$ associated with the size of the largest Q2D rectangular vortex loops.\cite{Chattopadhyay94,Shenoy95} We find that the RG flow is Q2D and dominated by spinwave excitations in the momentum range $l_J^{-1}<k<l_v^{-1}$. This gives rise to clear signatures of KT physics in a wide region of the low-temperature phase. We have shown that the order parameter $\rho_0$ is a tool of choice to identify the nature of the low-temperature phase. When $T\lesssim\Jperp$, $\rho_0$ varies quadratically with 
temperatures [Eq.~(\ref{rhosw1})] while for $\Jperp\lesssim T\lesssim\Tkt<T_c$ it varies as a power law of $\Jperp$ [Eq.~(\ref{rho0})] with a temperature-dependent exponent which depends on the anomalous dimension $\eta$ of the strictly 2D low-temperature KT phase. Furthermore the (square of the) order parameter is much smaller than the in-plane stiffness when the anisotropy is strong while both quantities are of the same order of magnitude in an isotropic system. Thus quasi-2D boson gases appear as a way to realize Bose-Einstein condensates with a highly depleted condensate density, which is unusual for weakly interacting superfluids. Our predictions could be tested with experiments on cold atoms where Q2D bosonic systems can be engineered with a tunable interplane coupling $\Jperp$.  

In the introduction we pointed out some limitations of the LPA$'$ so that one can wonder to what extent our results are robust with respect to the approximations made for solving the NPRG flow equations. In particular, algebraic order in the low-temperature phase is not obtained {\it stricto sensu} since the correlation length $\xi$ is always finite in the LPA$'$. Nevertheless, at low temperatures $\xi$ is so large that it cannot be seen in the RG flow even for running momentum scales as small at $10^{-15}\Lamb$. Moreover, the finite-temperature thermodynamics (e.g. the order parameter) is not sensitive to the deep infrared behavior of correlation functions. Whether we get a true line of fixed points with an infinite correlation length or only a line of quasi-fixed points with a large but finite correlation length is irrelevant for the thermodynamics. 

Our approach differs from most previous studies, based on the classical XY model, by including quantum fluctuations. The latter do not play an important role except for being responsible for the low-temperature boundary $T\simeq \Jperp$ below which the system behaves as a three-dimensional anisotropic system with a $T^2$ reduction of the order parameter wrt its zero-temperature value (the reduction would be linear in a classical model). Quantum fluctuations also contribute to the temperature dependence of the prefactor $\calA(T)$ in Eq.~(\ref{rho0}) and this may explain some differences with the numerical work of Ref.~\onlinecite{Furuya16} regarding the temperature dependence of the order parameter.

\begin{acknowledgments}
We thank N. Defenu, A. Trombenotti, N. Laflorencie, S. Capponi and E. Vicari for useful discussions and correspondence.
\end{acknowledgments}

\appendix

\begin{widetext}
\section*{Flow equations} 

Using the dimensionless variables $\tilde \q_\para=\q_\para/k$, $\tilde\w_n=\wn/c_{\para,k}k$ and 
\begin{equation}
\begin{split} 
&\tdelta_k = (Z_{\para,k}k^2)^{-1} \delta_k , \\ 
&\trho_{0,k} = k^{-1}(Z_{\para,k}Z_{\tau,k})^{1/2} \rho_{0,k} , \\
&\tlamb_k = k^{-1}Z_{\para,k}^{-3/2} Z_{\tau,k}^{-1/2} \lamb_k ,\\  
&\tilde Z_{\perp,k} = (Z_{\para,k}k^2)^{-1} Z_{\perp,k} ,
\end{split}
\end{equation}
we obtain the following RG equations 
\begin{equation}
\begin{split} 
\dt \tdelta_k &=  (\eta_{\para,k}-2)\tdelta_k + \frac{\tlamb_k}{2} ( 3\tilde I_{\rm L} +  \tilde I_{\rm T} ), \\ 
\dt \trho_{0,k} &= - \left(1+\frac{\eta_{\para,k}+\eta_{\tau,k}}{2} \right) \trho_{0,k} - \half ( 3\tilde I_{\rm L} + \tilde I_{\rm T} ) \\
\dt \tlamb_k &= \left( - 1+\frac{3\eta_{\para,k}}{2} +\frac{\eta_{\tau,k}}{2} \right) \tlamb_k - \tlamb_k^2 ( 9 \tilde J_{\rm LL} + \tilde J_{\rm TT} ) , \\ 
\eta_{\para,k} &= 2 \tlamb_k^2 \trho_{0,k} \partial_{\tilde \q^2} (\tilde J_{\rm LT} + \tilde J_{\rm TL} ) , \\ 
\eta_{\tau,k} &= 2 \tlamb_k^2 \trho_{0,k} \partial_{\twnu^2} (\tilde J_{\rm LT} + \tilde J_{\rm TL} ) , \\ 
\dt \tilde Z_{\perp,k} &= (\eta_{\para,k}-2) \tilde Z_{\perp,k} - 2 \tlamb_k^2 \trho_{0,k} \partial_{\pperp^2} (\tilde J_{\rm LT} + \tilde J_{\rm TL} ) ,   
\end{split}
\end{equation}
where $\eta_{\tau,k}=-\dt \ln Z_{\tau,k}$ ($\eta_{\para,k}$ and $\eta_{\perp,k}$ are defined by~(\ref{etadef})), 
with initial conditions 
\begin{equation}
\begin{split} 
U_\Lamb(\rho) &=  \left( r_0 - 2 \frac{J_\perp^2}{c^2} \right) \rho + \frac{u_0}{6} \rho^2 , \\ 
Z_{\para,\Lamb} &= 1, \quad Z_{\tau,\Lamb} = \frac{1}{c^2}, \quad Z_{\perp,\Lamb} = \frac{J_\perp^2}{c^2} .
\end{split} 
\end{equation} 
The threshold functions are defined by 
\begin{equation}
\begin{split}
\tilde I_\alpha ={}& \int_{\tilde\q_\para,\twn,\qperp} 
\frac{\eta_{\para,k} Yr+2Y^2r' + (r+Yr')[(\eta_{\tau,k}-\eta_{\para,k})\twn^2+(\eta_{\perp,k}-\eta_{\para,k})\tilde Z_{\perp,k}\eps_\perp]}{\tilde A_\alpha^2} , \\ 
\tilde J_{\alpha\beta} ={}&  \int_{\tilde\q_\para,\twn,\qperp} 
\frac{\eta_{\para,k} Yr+2Y^2r' + (r+Yr')[(\eta_{\tau,k}-\eta_{\para,k})\twn^2+(\eta_{\perp,k}-\eta_{\para,k})\tilde Z_{\perp,k}\eps_\perp]}{\tilde A_\alpha^2 \tilde A_\beta} , \\
\partial_{\tilde\q_\para^2} \tilde J_{\alpha\beta} ={}& -  \int_{\tilde\q_\para,\twn,\qperp} \tilde q_\para^2 \biggl\{ 
2 \bigl[\eta_{\para,k} Yr+2Y^2r' + (r+Yr')\bigl((\eta_{\tau,k}-\eta_{\para,k})\twn^2+(\eta_{\perp,k}-\eta_{\para,k})\tilde Z_{\perp,k}\eps_\perp\bigr)\bigr] \frac{\tilde A'_\alpha}{\tilde A_\alpha^3} \\ 
& - \bigl[ \eta_{\para,k} r + (\eta_{\para,k}+4)Yr' + 2Y^2r'' + (2r'+Yr'')\bigl((\eta_{\tau,k}-\eta_{\para,k})\twn^2+(\eta_{\perp,k}-\eta_{\para,k})\tilde Z_{\perp,k}\eps_\perp\bigr)\bigr] \frac{1}{\tilde A_\alpha^2} \biggr\} 
\frac{\tilde A'_\beta}{\tilde A^2_\beta} , \\ 
\partial_{\twnu^2} \tilde J_{\alpha\beta} ={}&  \int_{\tilde\q_\para,\twn,\qperp} \{\eta_{\para,k} Yr+2Y^2r' + (r+Yr')[(\eta_{\tau,k}-\eta_{\para,k})\twn^2 + (\eta_{\perp,k}-\eta_{\para,k})\tilde Z_{\perp,k}\eps_\perp]\} \frac{\tilde A_1^2-\tilde A_2 \tilde A_\beta}{\tilde A_\alpha^2 \tilde A_\beta^3} , \\
\partial_{\pperp^2} \tilde J_{\alpha\beta} ={}& -  \int_{\tilde\q_\para,\twn,\qperp} \frac{\eta_{\para,k} Yr+2Y^2r' + (r+Yr')[(\eta_{\tau,k}-\eta_{\para,k})\twn^2+(\eta_{\perp,k}-\eta_{\para,k})\tilde Z_{\perp,k}\eps_\perp]}{\tilde A^2_\alpha \tilde A^3_\beta} \\ & \times \biggl\{ \tilde A_\beta [ (1+r+Yr') \tilde Z_{\perp,k} \cos \qperp + 2 (2r'+Yr'') \tilde Z^2_{\perp,k} \sin^2 q_\perp] 
- 4 [ (1+r+Yr') \tilde Z_{\perp,k} \sin\qperp ]^2  \biggr\} 
\end{split}
\end{equation}
where 
\begin{equation}
\begin{split}
&\tilde A_{\rm T} = Y(1+r)+\tdelta_k + \tilde Z_{\perp,k} \eps_\perp(\qperp) ,  \qquad
\tilde A_{\rm L} = \tilde A_{\rm T} + 2\tlamb_k\trho_{0,k} , \qquad 
\tilde A'_\alpha = 1+r+Yr', \\
&\tilde A_1= 2\twn (1+r+Yr'), \qquad 
\tilde A_2=1+r+Yr' + 2\twn^2(2r'+Yr'') 
\end{split}
\end{equation}
and
\begin{equation}
\int_{\tilde\q_\para,\twn,\qperp} = \tilde T_k \sum_{\twn} \int \frac{d^2\tilde q_\para}{(2\pi)^2} \int_0^{2\pi} \frac{d\qperp}{2\pi} .
\end{equation}
\end{widetext}



%

\end{document}